\input{aipcheck}

\documentclass[
    ,final            
  ]
  {aipproc}

\layoutstyle{6x9}

\def\gsim{\lower0.5ex\hbox{$\:\buildrel >\over\sim\:$}}
\def\lsim{\lower0.5ex\hbox{$\:\buildrel <\over\sim\:$}}

\begin{document}

\title{Massive neutrinos in a Grounds-up Approach with Quark-Lepton Similarity}


\author{Shaouly Bar-Shalom}{
  address={Physics Department, Technion-Institute of Technology, Haifa 32000, Israel} ,altaddress={Speaker, E-mail: shaouly@physics.technion.ac.il}
}

\author{David Atwood}{
  address={Dept. of Physics and Astronomy, Iowa State University, Ames, IA 50011, USA}
}

\author{Amarjit Soni}{
  address={High Energy Theory Group, Brookhaven National Laboratory, Upton, NY 11973, USA}
}

\begin{abstract}
 We examine neutrino oscillations in a two Higgs
doublet model (2HDM) in which the second doublet couples only to
the third generation right-handed up-fermions, i.e., 
to $t_R$ and to $N_3$ which is the heaviest right-handed Majorana neutrino.
The inherently large $\tan\beta$ of this model
can naturally account for 
the large top mass and, based on a quark-lepton 
similarity ansatz, when embedded 
into a seesaw mechanism it can also account for the observed 
neutrino masses and mixing angles 
giving a very small $\theta_{13}$: 
$-0.017 \lsim \theta_{13} \lsim 0.021$ at 99\% CL, and a very 
restrictive 
prediction for the atmospheric mixing angle: 
$42.9^0 \lsim \theta_{atm} \lsim 45.2^0$ at 99\% CL.
The large value of $\tan\beta$ also sets the mass scale 
of the heaviest right-handed Majorana neutrino $N_3$ and 
triggers successful leptogenesis through a CP-asymmetry in the 
decays of the $N_1$ (lightest right-handed Majorana) 
which is $\tan^2\beta$ enhanced compared to 
the CP-asymmetry obtained in models for leptogenesis with one Higgs 
doublet or in the MSSM. 
This enhancement allows us to relax the lower bound on $M_{N_1}$ 
and consequently also the lower bound on the reheating temperature of 
the early universe.   
\end{abstract}

\maketitle


\section{Introduction}

  In the past decade we have witnessed two remarkable findings: 
(i) the discovery of the top-quark which turned out to be 
enormously heavy
compared to all the other fermions: 
$m_t \sim 175$ GeV, i.e., ``weighing'' almost as much as a Gold atom!, 
and (ii) the discovery of neutrino oscillations implying that neutrinos 
are massive with a typical mass in the sub-eV range, i.e., $m_\nu$ is 
more than 12 orders of magnitudes smaller than $m_t$.           
This two monumental discoveries of the 90's 
present us with the pressing challenge of reconciling the apparent 
enormous hierarchy in the masses of fundamental fermions.   

A possible resolution to this huge hierarchy between 
$m_\nu$ and $m_t$ may be encoded within the following triple-relation 
between $m_\nu$, $m_t$ (or the Electroweak scale) 
and the GUT mass-scale $M_{GUT} \sim 10^{16}$ GeV:
\begin{eqnarray}
m_\nu \sim \frac{m_t^2}{M_{GUT}} \label{relation}~.
\end{eqnarray}      
\noindent Indeed, the beautiful seesaw mechanism
dictates that (see also next sections):
\begin{eqnarray}
m_\nu \sim \frac{m_D^2}{M_{\nu_R}} \label{seesaweq}~,
\end{eqnarray}      
\noindent where $m_D$ is a Dirac neutrino mass term and 
$M_{\nu_R}$ is the mass of heavy right-handed Majorana neutrinos. 
Thus, based on the seesaw formula in Eq.~\ref{seesaweq}, 
the triple-relation in Eq.~\ref{relation} stands as a very strong 
hint for Dirac neutrino masses of $m_D \sim {\cal O}(m_t)$ 
and for the existence of super-heavy right-handed Majorana neutrinos 
with a typical mass of $M_{\nu_R} \sim {\cal O}(M_{GUT})$.  

In this work \cite{paper} we seriously take the triple-relation in 
Eq.~\ref{relation} at ``face-value'', suggesting that the 
impressive findings in the neutrino sector
are closely related to the heaviness of the top-quark.
In particular, we construct a model that,
based on a grounds-up approach, explicitly yields  
the triple-relation between the large $m_t$, the observed $\nu$-oscillation 
data (i.e., masses and mixing angles) and the super-heavy mass scale of the 
right-handed Majorana neutrinos. In addition, 
our model can rigger successful leptogenesis which 
can account for the observed Baryon asymmetry 
in the universe.
 
Our model is a two Higgs doublet model (2HDM)
which treats the 3rd generation
neutrino in a completely analogous manner to the top-quark.
We have, therefore,
named our model ``the 2HDM for the 3rd generation'' (3g2HDM).

\section{The two Higgs doublet model for the 3rd generation (3g2HDM)}

The 3g2HDM extends the 
idea of the so called ``2HDM for the top-quark'' (t2HDM) \cite{das}
to the leptonic sector. 
In particular, as in the t2HDM, we assume that $\phi_t$ [the 
Higgs doublet with a much larger vacuum expectation value (VEV)] couples 
{\it only} to 
the top-quark and to the 3rd generation right-handed Majorana neutrino, 
while the other Higgs doublet $\phi_f$ 
(with a much smaller VEV) couples to all the other fermions.
The large mass hierarchy between the top-quark
and all other fermions is then viewed as a consequence of
$v_t/v_f \equiv \tan\beta >> O(1)$,
which, therefore, becomes the ``working assumption'' of the 3g2HDM.

The Yukawa interaction Lagrangian of the 3g2HDM takes the form:
\begin{eqnarray}
{\cal L}_Y =
- Y^d {\bar Q}_L \phi_f d_R 
- Y_1^u {\bar Q}_L \tilde\phi_f u_R 
- Y_2^u {\bar Q}_L \tilde\phi_t u_R
- Y^e {\bar L}_L \phi_f \ell_R 
- Y_1^\nu {\bar L}_L \tilde\phi_f N 
- Y_2^\nu {\bar L}_L \tilde\phi_t N 
+h.c. \label{yukawa} ~,
\end{eqnarray}
\noindent where $N$ are right-handed Majorana neutrinos with a mass 
$M_N^{ij} N_i N_j/2$, $Q$ and 
$L$ are the usual quark and lepton doublets and the following Yukawa textures 
are assumed \cite{paper}
\begin{eqnarray}
Y_1^{u,\nu} \equiv \pmatrix{
a^{u,\nu} & b^{u,\nu} & 0  \cr
a^{u,\nu} & b^{u,\nu} & 0 \cr
0 & \delta b^{u,\nu} & 0 } ~,~
Y_2^{u,\nu} \equiv \pmatrix{
0 & 0 & 0  \cr
0 & 0 & c^{u,\nu} \cr
0 & 0 & c^{u,\nu} }
\label{yun}~,
\end{eqnarray}
\noindent such that, in both the quark and leptonic sectors,
$\phi_t$ couples only to the third generation right-handed 
up-fermions. Note also that  
$m_{D},~m_u = v_f (Y_1^{\nu,u} + \tan\beta Y_2^{\nu,u})/\sqrt{2}$, 
where $m_D,~m_u$ are the Dirac mass matrices of the neutrinos and 
up-quarks, respectively. 

\section{Neutrino oscillations in the 3g2HDM}

In the basis where $M_N$ is diagonal,
$M_N=M \cdot diag(\epsilon_{M1},\epsilon_{M2},\epsilon_{M3})$, we 
obtain from the seesaw mechanism formula $m_\nu = - m_D M_N^{-1} m_D^T$:
\begin{eqnarray}
m_\nu = 
m_\nu^0 \pmatrix{
\epsilon & \epsilon & \delta \bar\epsilon  \cr
\cdot & \epsilon+\omega & \delta \bar\epsilon+\omega \cr
\cdot & \cdot & \delta^2 \bar\epsilon+\omega } \label{mnu}~,
\end{eqnarray}
\noindent where 
\begin{eqnarray}
m_\nu^0 \equiv \frac{(v_1)^2}{2M} 
~,~ \epsilon \equiv \frac{a^2}{\epsilon_{M1}}+
\frac{b^2}{\epsilon_{M2}}~,~\bar \epsilon \equiv \epsilon -
\frac{a^2}{\epsilon_{M1}}~,~
\omega \equiv \frac{c^2 t_\beta^2}{\epsilon_{M3}} \label{omega}~.
\end{eqnarray}

\noindent In the following we will adopt a quark-lepton similarity 
Ansatz (perhaps motivated by 
GUT scenarios): $a^u \sim a^\nu \equiv a$, $b^u \sim b^\nu \equiv b$ and 
$c^u \sim c^\nu \equiv c$, with  
$a \sim O(10^{-3}),~b \sim O(10^{-1}),~c \sim O(1)$ which, in our model, 
follows from the up-quark sector since $a^u v_f \sim O(m_u)$, 
$b^u v_f \sim O(m_c)$ and $m_t \sim O(c^u v_f \tan\beta)$.
Then, diagonalizing the light-neutrinos mass matrix in 
Eq.~\ref{mnu},  
we find that in the normal mass-hierarchy 
scheme, i.e., $m_1 << m_2 << m_3$, \cite{paper}:
\begin{itemize}
\item The mass of the heaviest light-neutrino follows the triple relation 
in Eq.~\ref{relation}:
\begin{eqnarray}
m_{3} \sim \frac{m_t^2}{M_{N_3}} \label{relation2}~,
\end{eqnarray}  
\noindent where $M_{N_3} \sim M_{GUT}$ is the mass of the 3rd and heaviest 
right-handed Majorana neutrino.   
\item Performing a 
minimum $\chi^2$ analysis with respect to each of the 
oscillation parameters $\theta_{13},~\theta_{atm} \equiv \theta_{23},~
\theta_{sol}\equiv \theta_{12}$ and 
$\Delta m_{atm}^2$, $\Delta m_{sol}^2$, our 3g2HDM yields the following 
$99\%$ CL allowed ranges for the mixing parameters:
\begin{eqnarray}
28.0^0 \lsim &\theta_{sol}& \lsim 36.0^0 ~~~~99\% ~{\rm CL} ~, \nonumber \\
1.0 \cdot 10^{-3} ~ {\rm (eV)}^2 
\lsim &\Delta m_{atm}^2& \lsim 3.7 \cdot 10^{-3} ~ {\rm (eV)}^2 ~~~~99\% ~{\rm CL} ~, \nonumber \\
7.3 \cdot 10^{-5} ~ {\rm (eV)}^2 
\lsim &\Delta m_{sol}^2& \lsim 9.1 \cdot 10^{-5} ~ {\rm (eV)}^2 ~~~~99\% ~{\rm CL} ~, 
\end{eqnarray}  

\noindent with a very restrictive prediction for $\theta_{13}$ 
and the atmospheric mixing angle:
\begin{eqnarray}
-0.96^0 \lsim &\theta_{13}& \lsim 1.36^0 ~~~~99\% ~{\rm CL} ~, \nonumber \\ 
42.9^0 \lsim &\theta_{atm}& \lsim 45.2^0 ~~~~99\% ~{\rm CL} ~.
\end{eqnarray}  

\item The mass-spectrum of the heavy Majorana neutrinos (subject to the 
constraints coming from oscillation data) becomes:
\begin{eqnarray}
M_{N_3} \sim 100M~,~M_{N_2} \sim 0.01M~,~M_{N_1} >> 10^{-6}M  
\label{Mmassesa}~,
\end{eqnarray}
\noindent with $M \sim 10^{13}$ GeV.

\end{itemize}

\section{Leptogenesis in the 3g2HDM}

A CP-asymmetry, $\epsilon_{N_i}$, in the decay 
$N_i \to \ell \phi_j$ 
can generate the lepton asymmetry \cite{cpv}:
\begin{eqnarray}
n_L/s=\epsilon_{N_i} Y_{N_i}(T>>M_{N_i}) \eta ~,
\end{eqnarray}
\noindent where $Y_{N_i}(T>>M_{N_i})=135 \zeta(3)/(4\pi^4g_*)$
and $g_*$ being the effective number of spin-degrees of freedom in 
thermal equilibrium. Also,
$\eta$ is the ``washout'' parameter (efficiency factor) that measures
the amount of deviation from the out-of-equilibrium condition at 
the time of the $N_i$ decay.
This lepton asymmetry can then be converted into a baryon 
asymmetry through nonperturbative spheleron processes: In our case 
(i.e., two scalar doublets) we obtain:
\begin{eqnarray}
n_B/s \sim -1.4\times 10^{-3} \epsilon_{N_i} \eta \label{nb}~.
\end{eqnarray}

\noindent As seen from Eq.~\ref{Mmassesa}, 
our 3g2HDM can lead 
to a hierarchical mass spectrum for the heavy Majorana neutrinos, 
$M_{N_1}<<M_{N_2}<<M_{N_3}$.
In this case, only the CP-asymmetry produced 
by the decay of $N_1$ survives, i.e., $\epsilon_{N_i} \to \epsilon_{N_1}$ 
and we get:

 \begin{eqnarray}
\epsilon_{N_1} \sim 
- \frac{3}{16 \pi} 
\frac{t_\beta^2 \sqrt{{\Delta m}_{sol}^2}}{m_t^2} 
\epsilon M_{N_1} \sin 2(\theta_b-\theta_a) 
\sim - \epsilon_{N_1}^{max} \times \frac{2 \epsilon}{\omega} t_\beta^2  \sin 2(\theta_b-\theta_a) \label{eps}~,
\end{eqnarray}

\noindent where the CP-phases arise from the possible complex 
entries in $Y_1^\nu$: 
$a=|a|e^{i\theta_a}$ and $b=|b|e^{i\theta_b}$, and  
$\epsilon_{N_1}^{max}$ is the 
maximum of the CP-asymmetry in models where only one Higgs doublet 
couples to the neutrino fields (see e.g., \cite{cpv}).
Thus, the CP-asymmetry in the 3g2HDM 
is a factor of $ \sim \frac{2 \epsilon}{\omega} \times t_\beta^2 \sim 20$ larger than the CP-asymmetry in the SM or the in MSSM, since 
in our model $\epsilon \sim 0.5$ and $\omega \sim 5$ are  
fixed by oscillation data and $t_\beta \sim {\cal O}(10)$ in order 
to account for the large top-quark mass. 

For the Baryon asymmetry we then obtain (by calculating the washout factor $\eta$ and using Eq.~\ref{nb} \cite{paper}): 
\begin{eqnarray}
\frac{n_B}{s} \sim 10^{-17} \tan^2\beta 
\frac{ \sqrt{{\Delta m}_{sol}^2}}{2 m_t^2} 
\epsilon M_{N_1} \left(\frac{M_{N_1}}{{\rm GeV}}\right)^{1.2} 
\sin 2(\theta_b-\theta_a) \label{basym}~. 
\end{eqnarray}

\noindent 
Eq.~\ref{basym} has to be compared with the observed baryon to photon 
number ratio 
$n_B/n_\gamma \sim 6 \times 10^{-10}$, implying 
$n_B/s \sim 8.5 \times 10^{-11}$.
For example, taking $\epsilon \sim 0.5$ and 
${\Delta m}_{sol}^2 \sim 8.2 \cdot 10^{-5}$ eV$^2$ (these values are consistent with the observed oscillation data), along with
$t_\beta \sim 10$ and $m_t \sim 170$ GeV, Eq.~\ref{basym} 
reproduces the observed baryon asymmetry for e.g., 
$M_{N_1}\sim 10^{10}$ GeV and $\sin 2(\theta_b-\theta_a) \sim 0.1$, or
for $M_{N_1}\sim 10^{9}$ if CP is maximally violated in the sense 
that $\sin 2(\theta_b-\theta_a) \sim 1$.  




\end{document}